# Metropolitan-scale Entanglement Distribution with Co-existing Quantum and Classical Signals in a single fiber.

A. Rahmouni[1,*], P. S. Kuo[1], Y.S. Li-Baboud[1], I. A. Burenkov[1,2], Y. Shi[1], M. V. Jabir[1], N. Lal[1], D. Reddy[3], M. Merzouki[1], L. Ma[1], A. Battou[1], S. V. Polyakov[1,2], O. Slattery[1,*], T. Gerrits[1,*]

[1]National Institute of Standards and Technology, 100 Bureau Drive, Gaithersburg, MD 20899, USA
[2]Physics Department, University of Maryland, College Park, MD 20742, USA
[3]National Institute of Standards and Technology, 325 Broadway, Boulder, CO 80305, USA
*Corresponding author: anouar.rahmouni@nist.gov; oliver.slattery@nist.gov; thomas.gerrits@nist.gov

The development of prototype metropolitan-scale quantum networks is underway and entails transmitting quantum information via single photons through deployed optical fibers spanning several tens of kilometers. The major challenges in building metropolitan-scale quantum networks are compensation of polarization mode dispersion, high-precision clock synchronization, and compensation for cumulative transmission time fluctuations. One approach addressing these challenges is to co-propagate classical probe signals in the same fiber as the quantum signal. Thus, both signals experience the same conditions, and the changes of the fiber can therefore be monitored and compensated. Here, we demonstrate the distribution of polarization entangled quantum signals co-propagating with the White Rabbit Precision Time Protocol (WR-PTP) classical signals in the same single-core fiber strand at metropolitan-scale distances. Our results demonstrate the feasibility of this quantum-classical coexistence by achieving high-fidelity entanglement distribution between nodes separated by 100 km of optical fiber. This advancement is a significant step towards the practical implementation of robust and efficient metropolitan-scale quantum networks.

## 1. INTRODUCTION

Quantum mechanics promises the unprecedented ability to secure communications across deployed network infrastructures with security guarantees unattainable with classical systems [1-3]. This evolving frontier of research foresees a future where quantum information can be securely distributed and shared among quantum computers [4], clusters of quantum sensors [5], and related devices spanning local, regional, national, and global distances [6]. The establishment of metropolitan-scale quantum networks is a crucial milestone in realizing the benefits of secure distributed communications and sensing. Currently, significant progress is being made in the development of metropolitan-scale prototype quantum networks in key cities such as New York, Chicago, Boston, Washington DC, as well as others across the globe [7-12]. As an example, the Washington Metropolitan Quantum Network Research Consortium (DC-QNet), aims to develop, demonstrate, and operate a quantum network serving as a regional testbed, based on entanglement distribution [13].

Among the major challenges in metropolitan-scale quantum networks, which entails transmitting quantum information via photons through deployed optical fibers spanning several tens of kilometers, compensation for polarization mode dispersion, timing for node synchronization, and compensation for cumulative transmission time fluctuations are most notable [14, 15]. These challenges are particularly acute when implementing advanced protocols such as entanglement swapping. Co-existence of classical and quantum signals in the same fiber represents an effective approach to address these challenges. In co-existence, classical probe signals are transmitted in the same fiber as the quantum signals [16] and will therefore experience the same transmission conditions as the quantum signals. The classical probe signal can be characterized to provide the relevant telemetry data that can be used to compensate for varying fiber conditions in real-time. Moreover, the scalability of quantum networking will gain from the coexistence of quantum and classical communications in shared fibers. This is especially significant given the limited availability of fibers in practical metropolitan networks.

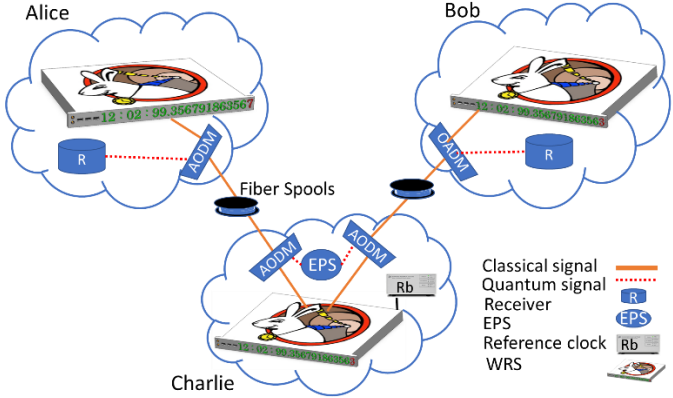

Figure 1: Working principle of entanglement distribution between two distant nodes (Alice and Bob) co-existing with classical time synchronization using the White Rabbit precision time protocol (WR-PTP). The classical signal is a bidirectional optical communication between the White Rabbit Switches (WRSs). The quantum signal is generated by a polarization-entangled photon source (EPS) at Charlie, with the first photon sent to Alice and the second photon sent to Bob. The optical add-drop multiplexers (OADMs), which contain a series of CWDMs and DWDMs, are used to multiplex and demultiplex the WR-PTP signal and the quantum signal. See [17] for WR logo copyright information.

In addition, high-accuracy synchronization plays a fundamental role in many quantum network architectures [18]. Because single-photon wavepackets in some qubit systems can be as short as picoseconds, high-accuracy synchronization over long distances in deployed fibers is required [19, 20]. For example, when using spontaneous parametric down conversion (SPDC) as a source for entanglement distribution and swapping, picosecond synchronization will be required between the nodes [21]. However, achieving such precise synchronization in long, deployed fibers poses challenges due to environmental conditions and weather-induced fluctuations [22]. Notably, cumulative transmission time fluctuations increase significantly due to variations in the refractive index and physical length of the fiber. To probe the timing fluctuations within the fiber used for transmission of quantum information, a synchronization signal should propagate within the same fiber. The White Rabbit Precision Time Protocol (WR-PTP), which is now part of the IEEE 1588-2019 High Accuracy Precision Time Protocol[1] has been employed for clock synchronization in quantum network testbed implementations [21, 23, 24]. WR-PTP is a classical optical-two-way time transfer protocol to provides sub-nanosecond synchronization accuracy and picosecond-scale precision well-suited for quantum network implementation.

Here, we demonstrate the co-existence of the WR-PTP with the quantum signal and therefore offer a solution to compensate for cumulative transmission time fluctuations in the same fiber as the quantum signal.

A prominent challenge in co-propagating quantum signals in the same fiber strand as the classical signals is the considerable amount of broadband noise generated by the high optical power of classical signals. This noise is particularly significant on the Stokes side [25]. Previous research studied the co-existence of classical and quantum signals, using the O-band for quantum signals and the C-band for classical signals, to showcase its feasibility over metropolitan-scale distances [26-28]. On the contrary, the exploration of employing C-band for quantum signal alongside O- or L-band optical clock distribution systems in a previous study conducted over 59 km of deployed fiber between Fermi and Argonne National Laboratories, resulted in a reduction of the Coincidence-to-Accidental Ratio (CAR) from $51 \pm 2$ to $5.3 \pm 0.4$ [29]. Despite this progress, to the best of our knowledge, achieving entanglement distribution with such co-existence over longer metropolitan-scale distances exceeding 100 km has not previously been demonstrated. Furthermore, the configuration of using the C-band for quantum signals and the O-band for classical signals benefits from capitalizing on the low-loss regime in optical fibers. The difference is approximately ~0.1 dB/km, equivalent to 10 dB for 100 km. This configuration also takes advantage of existing telecom infrastructure, such as Wavelength Division Multiplexing (WDM) systems, especially crucial in the context of limited fiber availability in real-world metropolitan networks. In this work, we demonstrate the implementation of polarization entanglement distribution in the C-band over metropolitan-scale distances in which the quantum signal is coexisting with an O-band WR-PTP signal on a single fiber. Measurements were conducted over approximately 250 meters of deployed fiber across the NIST-Gaithersburg campus, with an extension to over 100 km using fiber spools.

To achieve co-existence over such a long distance, we incorporated small form-factor pluggable (SFP) modules with a receiver sensitivity of < -41 dBm. The chosen wavelengths for transmitting (Tx) and receiving (Rx) signals were carefully selected to be below 1300 nm (specifically at 1290 nm and 1270 nm for Tx and Rx, respectively), aimed to minimize Raman scattering noise [4]. Additional measurement results of the CAR as a function of fiber length emphasize the practicality of this configuration, showcasing an impressive reach of up to 120 km between two separated nodes. This technological advance holds great promise for achieving the co-existence of classical and quantum signals of deployed metropolitan-scale quantum networks.

## 2. EXPERIMENTAL CONFIGURATION

Figure 1 illustrates the working principle of entanglement distribution between two distant nodes where the quantum signal co-exist with the WR-PTP classical signal. In this configuration, three WR-PTP switches were employed, with the grandmaster (GM) disciplined by a Rubidium (Rb) reference clock located at Charlie's node. The (SPDC) based entangled photon source (EPS) is also situated at Charlie's node. The quantum and classical signals co-propagate across the fiber to both, Alice's and Bob's node, where the receivers and WR-PTP boundary clocks (BCs) are located. The classical signal involves bidirectional optical communication for implementing the WR-PTP using the SFP modules. The OADMs, which contain a series of coarse WDMs (CWDMs) and dense WDMs (DWDMs), are used to multiplex and demultiplex the WR-PTP signal and the quantum signal. The receiver includes a polarization analyzer and detection components as shown.

The experiment is performed between two buildings on the Gaithersburg campus of the National Institute of Standards and Technology (NIST), as depicted in Fig. 2 (a). Alice's node is located at the end of building 221, Charlie's node is located at the other end of the same building, and Bob's node is located in the adjacent building 222. Charlie's node hosts the source (middle pane in Figure 2(b) for generating the polarization entangled photons in the telecom C-band. We built the SPDC source using type II based on a commercially available periodically-poled potassium titanyl phosphate (PPKTP) waveguide. The SPDC process is pumped by a narrowband continuous-wave (CW) laser at 778.6 nm. The signal and idler photons are coupled into a PANDA[1] polarization maintaining fiber (PMF), oriented such that the fiber's fast axis is rotated by 90° compared to that of the waveguide to compensate for the temporal walk-off between the $|H\rangle$ and $|V\rangle$ photons that is caused by the waveguide birefringence [30, 31]. The correct compensation length of the PMF is 1.18 m, measured by using a two-photon interference experiment based on a Hong-Ou-Mandel (HOM) interference [31]. We then use a 2-channel DWDM with a 100 GHz channel separation to filter the photons into the ITU-24 (1558.17 nm) and ITU-25 (1557.36 nm) channels, respectively. These two channels satisfy energy conservation with the pump and enable creation of a polarization-entangled state [32].

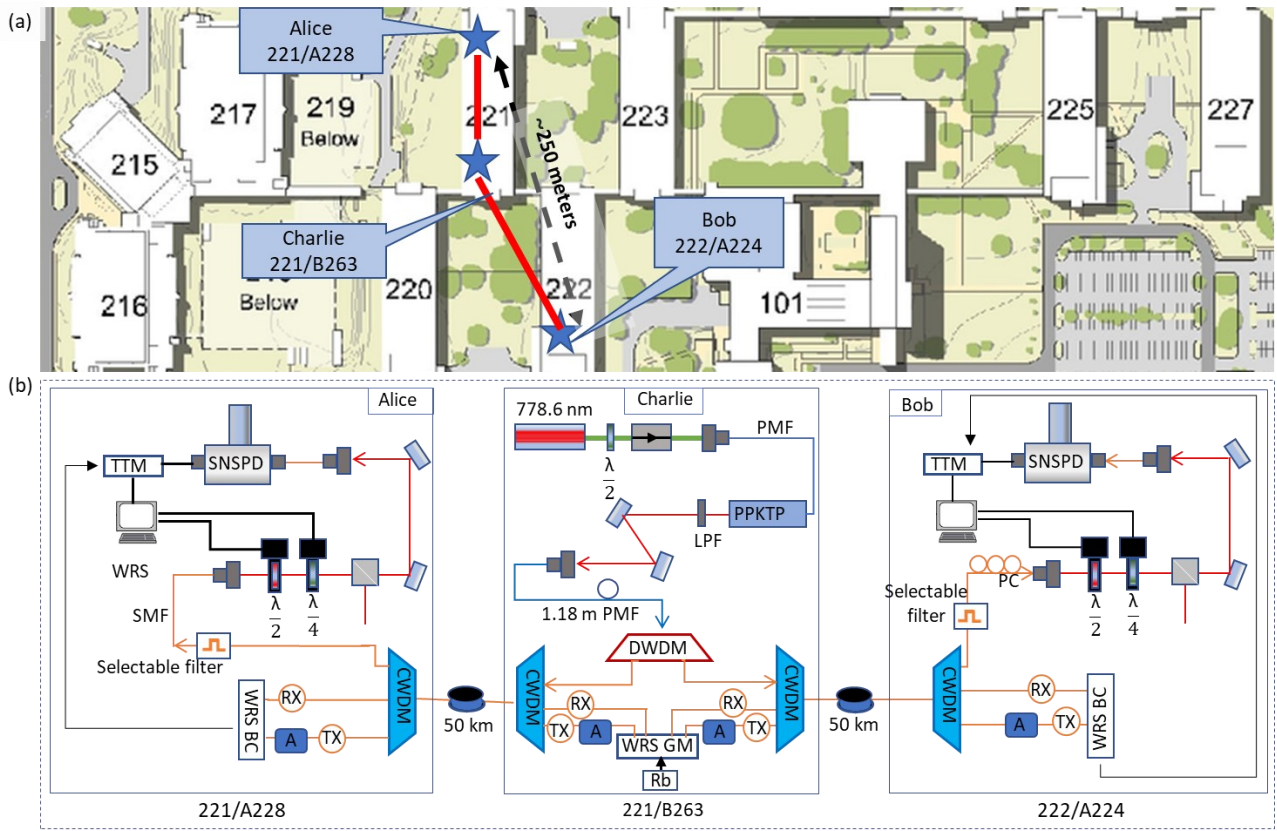

Figure 2: Experimental implementation of long-distance entanglement distribution using a single fiber co-existing quantum signals and classical synchronization signals (a) Optical path schematic and distance mapping: a visual representation of the optical path between two NIST buildings is depicted, illustrating 250 meters of deployment of fiber between Alice and Bob nodes. (b) Node configurations: detailed configurations at each node are presented, showcasing the EPS located at Charlie between Alice and Bob. Fiber spools are incorporated at each receiver node (Alice and Bob) to extend the fiber length. Charlie emits polarization entangled state photons to receivers positioned at Alice and Bob. The synchronization of receiver stations at both nodes is achieved using the WRPTP architecture, where the WRPTP classical signal and the quantum signal co-exist within the same fiber. Key components include superconducting nanowire single photon detectors (SNSPD), single-mode fiber (SMF), Time-tagging module (TTM), White Rabbit Switch (WRS), polarization maintaining fiber (PMF), long-pass filter (LPF), polarization controller (PC), and Rubidium clock (Rb).

Each receiver consists of a polarization-entanglement analyzer and a superconducting nanowire single-photon detector (SNSPD). The setup for each of the analyzers (at Alice and Bob) shown in figure 2(b) includes a motorized quarter-wave plate (QWP), a motorized half-wave plate (HWP), and a polarizing beam-splitter (PBS) [33]. We used a polarization controller (PC) at Bob's receiver to carefully align the polarization at the output of the fiber with respect to Alice's analyzer. Two SNSPDs were used for the detection of the photons at Alice and Bob with a detection efficiency >85 % at 1550 nm and a dark count rate of less than 500 counts/s. The distance between the nodes is extended using various combinations of optical fiber spools of 25 km and 50 km in length.

The two WRSs at Alice's and Bob's nodes are used to synchronize and calibrate the time base of the time-tagging modules (TTMs). The WRSs transmit the WR-PTP signals at 1270 nm and 1290 nm, respectively, using 1 Gbps Ethernet in the same fiber as the quantum signal generated from the entangled photon source. The CWDMs are used to multiplex and demultiplex the WR-PTP signal and the quantum signal. The selectable DWDM filters are used to reduce the noise and pass only photons in the quantum channel's frequency band to the receiver. The measurements are performed using TTMs, which record the arrival times of the photons based on the WRS time reference for each detection event. Bob sends his recorded time-tag data to Alice via the classical ethernet protocol, in real-time, where Alice analyses the coincidence counts.

Additional measurements were carried out to analyze the noise produced in the presence of classical-signal co-existence. The characterization involved measuring the CAR as a function of both the fiber length and the coincidence timing window. To simplify the experiment, particularly when not measuring entanglement fringes, we adopted a strategy of transmitting the two quantum signals (signal and idler) in the same direction from the source (at Charlie's node) to the analyzer (Alice's node) in coexistence with the WR-PTP classical signal; that is, all signals propagate in the same fiber between Alice and Charlie. Figure 3 illustrates the experimental setup used for this measurement. The signal (1557.36 nm) propagation path toward Alice remains with the same as that of the entanglement-distribution setup, except that the DWDM, previously used at the source (Charlie's node) to separate the signal and idler, is now moved to Alice. This modification resulted in the same noise as observed in the previous experiment. On the other side, the idler (1558.17 nm) co-propagates alongside the signal. The resulting noise due to Raman scattering is equivalent to sending signal and idler photons across two different fibers. The total transmitted distance between signal and idler is therefore twice the fiber length. The distance between the source (Charlie) and Alice's node is approximately 100 meters, plus the length of the extended fiber spools.

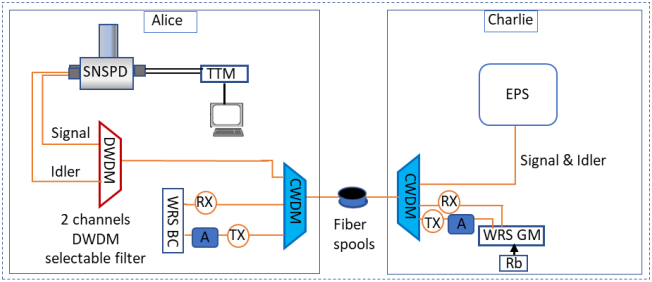

Figure 3: Configuration setup for noise characterization. The Entangled Photon Source (EPS) is depicted in Figure 2. The distance between the source (Charlie) and the Alice's node is approximately 100 meters, plus the length of the extended fiber spools. The signal and idler are separated after traversing the fiber at Alice. Two SNSPD detectors are employed to detect signal and idler photons, with their electrical outputs connected to a single Time Tagging Module (TTM) for coincidences measurement.

## 3. RESULTS AND DISCUSSIONS

To quantify the quality of the entangled state after co-propagating with the classical signal we extract the Clauser, Horne, Shimony and Holt (CHSH) S parameter [9] from interference fringe measurements for 250 meters (Figure 4(a)), 50 km (Figure 4(b)) and 100 km (Figure 4(c)) co-propagation distance. The entanglement fringes are obtained in the four measurement basis states ($-\frac{\pi}{4}$, 0, $+\frac{\pi}{4}$ and $+\frac{\pi}{2}$) by rotating Bob' polarizer angle. The datapoints in figure 4 correspond to experimental raw counts and the solid lines correspond to a fit of the data to a sine-square curve. Using a 2 ns coincidence window, we reached maximum coincidences rates of approximately 5100, 275 and 170 counts per second for 250 meters, 50 km and 100 km fiber length between Alice and Bob, respectively.

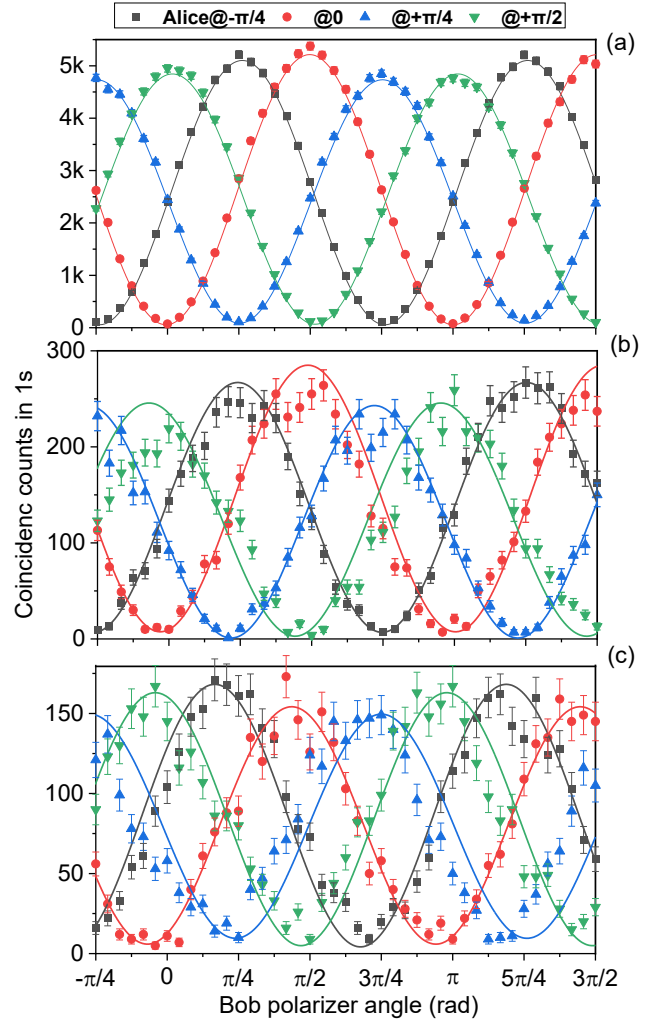

Figure 4: Polarization entanglement fringes as measured with 1 s integration time and co-existing with the WRPTP signal on varies fiber lengths: (a) deployed 250 meters fiber across the NIST campus, (b) 50 km extended fiber spool. (c) 100 km extended fiber spool.

Table 1 shows the entanglement fringe visibilities for the four basis states, calculated using the formula $V=(N_{max}-N_{min})/(N_{max}+N_{min})$, where $N_{max}$ and $N_{min}$ represent the maximum and minimum coincidence counts rates obtained from the fitted curves. Our results reveal entanglement fringe visibilities > 94% for entanglement distribution over a short distance of 250 meters across deployed fiber. When extending the distance using fiber spools to 50 km (25 km spools on each receiver side), the extracted visibilities remain consistently above 90%. Further extension to over 100 km of fiber spools (50 km spools on each receiver side) yielded estimated visibilities exceeding 81%, even without background noise subtraction. These results surpass the theoretical limit of $1/\sqrt{2} \approx 71\%$ required to confirm entanglement of photon pairs through Bell's test.

Moreover, the derived CHSH inequality parameter (S) [34, 35], calculated from coincidence-count fringes in the four basis states, serves as a robust indicator of the quality of entanglement. Fringes exhibiting S values greater than 2 signify violation of Bell's inequalities and cannot be explained classically. In our experimental results, the extracted S values are significantly non-classical, registering at 2.78 ±0.02, 2.64±0.03, and 2.30±0.04 for distances of 250 meters, 50 km, and 100 km, respectively.

**Table 1: Visibility and S parameter values extracted from the fitted curves of the measured fringe results, see figure 4.**

|  | Basis states | 250 Meters | 50 Km | 100 Km |
|---|---|---|---|---|
| Visibility | $-\frac{\pi}{4}$ | 0.96+0.02 | 0.90+0.01 | 0.87+0.02 |
|  | 0 | 0.96+0.02 | 0.92+0.01 | 0.86+0.02 |
|  | $+\frac{\pi}{4}$ | 0.94+0.03 | 0.91+0.02 | 0.81+0.03 |
|  | $+\frac{\pi}{2}$ | 0.95+0.02 | 0.90+0.02 | 0.81.0.02 |
| CHSH (S) |  | 2.78 ±0.02 | 2.64±0.03 | 2.30±0.04 |

Quantum state tomography measurements allow us to estimate the full state of the distributed quantum state [35, 36]. We reconstruct density matrices for the distributed quantum states, as graphically represented in Figure 5, from which we calculate the state's fidelities. The average quantum-state fidelity for 50 km and 100 km co-propagation distances are 0.96 and 0.87, respectively.

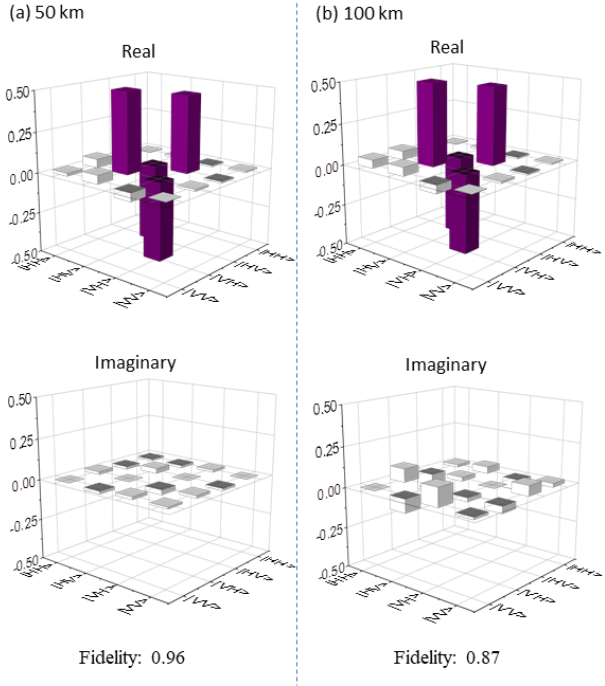

Figure 5: Graphical representation of the density matrix estimated by a quantum state tomography measurement. The top plots are the real parts and the bottom plots are the imaginary parts. Quantum state tomography results for the entangled state measured over (a) 50 km fiber length and (b) over a 100 km fiber length.

The predominant factor impacting the quality of entanglement within the configuration of co-existing quantum and classical signals in the same fiber is the additive noise due to Raman-scattered photons arising from the large optical power of the classical signals, especially when the quantum signal is situated on the Stokes side. This is the configuration employed here where the classical light is in the O-band and the quantum signal is in the C-band. However, opting for the C-band for the quantum signal may present a strategic advantage, leveraging existing telecom infrastructure, including DWDM systems, and capitalizing on the minimal optical fiber attenuation around 1550 nm.

The two indispensable parameters crucial for achieving such an extended distance co-existing with classical signal are the receiver's sensitivity and the transmission wavelengths. Initially, different SFP modules were used with a receiver sensitivity of -17 dBm, and the bi-directional transmission (Tx and Rx) wavelengths between the WRS leader and followers were set at 1310 nm and 1290 nm, respectively. The maximum distance attained between the two receivers was 25 km as detailed in Table 2. Subsequently, we incorporated SFP modules with an enhanced receiver sensitivity of < -41 dBm and selected the new Tx and Rx wavelengths of 1290 nm and 1270 nm, respectively. Notably, the use of 1270 nm allowed for a more than tenfold reduction in Raman scattering noise compared to using 1310 nm [16]. We point out that classical receiver sensitivity can be significantly improved, e.g. by taking advantage of the quantum measurement [37], in which case significantly longer links can be implemented. However, the goal of this work is to use off-the shelf components.

**Table 2: Comparison of visibility measurements using two pairs of SFP modules, one pair using upstream and downstream transmission wavelengths at 1290 nm and 1310 nm with -17 dBm sensitivity, and the other pair with upstream and downstream transmission wavelengths at 1270 nm and 1290 nm with -41 dBm sensitivity.**

| SFP wavelengths (nm) | SFP sensitivity (dBm) | 250 meter | 25 km | 50 km | 100 km |
|---|---|---|---|---|---|
| 1310/1290 | -17 | 93 % | 85% | NA | NA |
| 1290/1270 | -41 | 96 % | - | 90 % | 87 % |

To delve deeper into the impact of noise photons generated across the fiber and its dependence on the fiber length and the coincidence-window width, we conducted additional optical characterization measurements using the experimental setup shown in Figure 3. In this experiment, we measured the noise and the CAR for different coincidence windows as a function of fiber length, conducted prior to the entanglement polarization analyzer. The Raman scattered noise measured for 50 km and 100 km are 8500 and 32600 counts per second, respectively. As depicted in Figure 6, the Raman scattered noise increases with optical fiber length. According to the exponential fitted curve (red solid curve), the noise is estimated to increase dramatically for distances greater than 120 km, reaching over 1.5 million counts per second for a 200 km distance. Conversely, the CAR decreases with an increase in optical fiber length. The CAR remains above 10 for distances of up to 120 km, sufficient to achieve a visibility exceeding the 71 percent threshold required for violating Bell's inequalities.

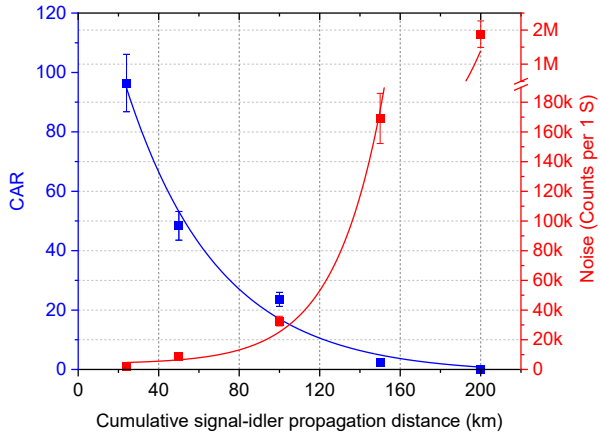

Figure 6: Measured coincidence-to-accidental ratio (CAR, left y axis) and the detected noise count rate (right y axis) as a function of length-of-flight across fiber (2x fiber length) with 2 ns of timing window. Note the broken right y-axis used to display the significantly elevated value of the measured noise count rate at 200 km.

The coincidence-window width stands as a critical parameter influencing the CAR. A narrower time window has the advantage of reducing accidental counts, thereby reducing the noise. However, it may also lead to a reduction in the coincidence counts rates, and vice versa. Achieving an optimal balance between CAR and total coincidence rates necessitates careful optimization of the coincidence timing window, tailored to the specifics of the optical setup, particularly depending on the characteristics of the EPS.

In this context, we conducted measurements of coincidence counts and CAR using a total fiber length of 50 km. This involved co-propagating the signal and idler within the same fiber, resulting in a total propagation distance of 100 km—an equivalent scenario to transmitting signals between two distant nodes as described above. The analysis focused on observing the dependence of CAR and coincidences as a function of the coincidence window width, as illustrated in Figure 7. We observed that CAR values show a plateau at approximately a 1 ns window width before progressively decreasing for widths larger than 1 ns. For the same reason, the coincidence counts increase progressively for increasing widths until about 1 ns after which the detection rates reach a plateau.

In our study, we used 2 ns coincidence window width to ensure all coincidences were recorded. However, upon studying the effect of the coincidence window on the CAR, we found that a shorter window would have resulted in better CAR. Therefore, with such potential improvement in the CAR, which further supports the feasibility of achieving entanglement distribution over 120 km, especially with further optimization efforts, such as optimizing the coincidence window, selecting the classical wavelength further away from the quantum signal, fine-tuning the power of the classical signal, and selecting filters with low insertion loss and optimized bandwidth. Despite that, we successfully implemented polarization entanglement distribution with the quantum signal co-existing with WR-PTP on a single fiber link spanning over 100 km.

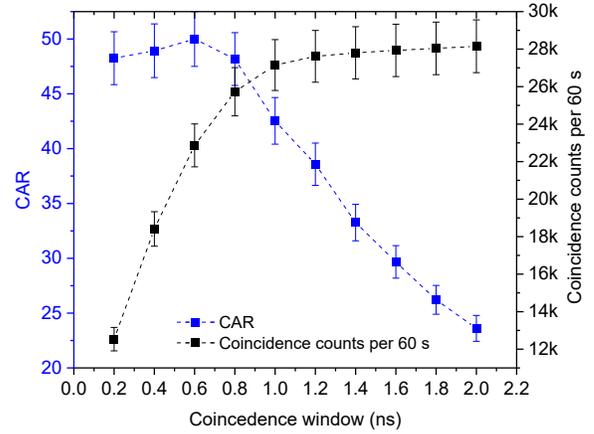

Figure 7: Measured CAR (left y axis) and the detected coincidence count rate (right y axis) as a function of coincidence timing window in co-existent experiment with the length-of-flight of 100 km.

## 4. Conclusion

We have successfully demonstrated the distribution of polarization entanglement in which the quantum signals are co-existing with the WR-PTP classical signal over the same fiber, spanning approximately 250 meter of locally deployed fiber and extended to over 100 km using fiber spools. We selected the C-band for quantum signals and the O-band for classical signals, leveraging existing telecom infrastructure, including DWDMs, and capitalizing on low signal loss around 1550 nm. The achieved entanglement fringe visibility surpassed 87%, resulting in a CHSH inequality parameter of 2.30. Quantum state tomography confirmed a fidelity of 0.87, attesting to the robustness of the entanglement distribution. To our knowledge, this is the first demonstration of polarization entanglement distribution over metropolitan-scale distances where the quantum and classical signals share the same fiber.

Furthermore, our thorough characterization substantiates the potential of this configuration to cover distances exceeding 120 km, while maintaining high entanglement visibility. Our experimental demonstration of entanglement distribution with co-existing classical and quantum signals marks a significant advance in the field of quantum networking. With a limited supply of available fibers in real-world metropolitan networks, our results offer a promising solution for the implementation quantum networking where the classical header information (e.g., probe-signals, router) propagate alongside quantum traffic in the same fiber.

**Disclaimer:** [1]Certain commercial equipment, instruments, or materials are identified in this paper to foster understanding. Such identification does not imply recommendation or endorsement by the National Institute of Standards and Technology, nor does it imply that the materials or equipment identified are necessarily the best available for the purpose.